\documentclass{elsart}

\usepackage{epsfig}
\usepackage{amssymb}
\usepackage{graphicx}
\usepackage{dcolumn}
\newcommand{\beq}{\begin{equation}}
\newcommand{\eeq}{\end{equation}}
\newcommand{\beqar}{\begin{eqnarray}}
\newcommand{\eeqar}{\end{eqnarray}}
\newcommand{\ds}{\displaystyle}
\begin{document}
\begin{frontmatter}

\title{ Elliptic flow at RHIC: where and when does it formed? }

\author[oslo,msu]{L.V. Bravina}
\author[oslo]{K. Tywoniuk}
\author[oslo,msu]{E.E. Zabrodin}
\author[tueb]{G. Burau}
\author[tueb]{J. Bleibel}
\author[tueb]{C. Fuchs}
\author[tueb]{Amand Faessler}
\address[oslo]{Department of Physics, University of Oslo,
Blindern 1048, N-0316 Oslo, Norway}
\address[msu]{Skobeltzyn Institute for Nuclear Physics, Moscow State
University, Vorobievy Gory, RU-119899 Moscow, Russia}
\address[tueb]{Institute for Theoretical Physics, University of
T\"ubingen, Auf der Morgenstelle 14, D-72076 T\"ubingen, Germany}


\begin{abstract}
Evolution of the elliptic flow of hadrons in heavy-ion collisions at 
RHIC energies is studied within the microscopic quark-gluon string
model. The elliptic flow is shown to have a multi-component structure
caused by (i) rescattering and (ii) absorption processes in spatially
asymmetric medium. Together with different freeze-out dynamics of
mesons and baryons, these processes lead to the following trend in 
the flow formation: the later the mesons are frozen, the weaker their
elliptic flow, whereas baryon fraction develops stronger elliptic flow
during the late stages of the fireball evolution.
Comparison with the PHOBOS data demonstrates the model ability to
reproduce the $v_2^{ch} (\eta)$ signal in different centrality bins.
\end{abstract}

\begin{keyword}
ultrarelativistic heavy-ion collisions \sep elliptic flow \sep 
freeze-out of particles \sep Monte-Carlo quark-gluon string model  

\PACS 25.75.-q \sep 25.75.Ld \sep 24.10.Lx 
\end{keyword}
\end{frontmatter}

%

\section{Introduction}
\label{sec1}

Elliptic flow is defined as the second harmonic coefficient $v_2$ 
of an azimuthal Fourier expansion of the particle invariant 
distribution \cite{VZ96}
\beq
\ds
E \frac{d^3 N}{d^3 p} = \frac{1}{\pi} \frac{d^2 N}{dp_t^2 dy} \left[
1 + 2 v_1 \cos(\phi) + 2 v_2 \cos(2 \phi) + \ldots \right] \ ,
\label{eq1}
\eeq
where $\phi$ is the azimuthal angle between the transverse momentum of 
the particle and the reaction plane, and $p_t$ and $y$ is the 
transverse momentum and the rapidity, respectively. The first harmonic
coefficient $v_1$ is called directed flow. It can be presented as
\beq \ds
v_1 \equiv \langle \cos{\phi} \rangle = 
\left \langle \frac{p_x}{p_t} \right \rangle \ , 
\label{eq2}
\eeq
while the $v_2$, which measures the eccentricity of the particle 
distribution in the momentum space, is
\beq 
\ds
v_2 \equiv \langle \cos{2 \phi} \rangle = 
\left \langle \frac{p_x^2 -p_y^2}{p_t^2} \right \rangle \ .
\label{eq3} 
\eeq 
(Note, that in the coordinate system applied the $z$-axis is directed
along the beam, and the impact parameter axis is labeled as $x$-axis.
Therefore, transverse momentum of a particle is simply 
$p_t = \sqrt{p_x^2 + p_y^2}$ ).

The overlapping area of two nuclei colliding with non-zero impact
parameter $b$ has a characteristic almond shape \cite{Sor97} in the
transverse plane. The fireball tries to restore spherical shape, 
provided the thermalization sets in rapidly and the hydrodynamic
description is appropriate \cite{Olli92,PoVo98,KSH00,TLS01}. When it 
becomes spherical, apparently, the elliptic flow stops to develop. 
Therefore, $v_2$ can carry important information about the earlier 
phase of ultrarelativistic heavy-ion collisions, equation of state 
(EOS) of hot and dense hadronic (or rather partonic) matter, and is 
expected to be a useful tool to probe the formation and hadronization 
of the quark-gluon plasma (QGP) \cite{QM02}.

Elliptic flow of charged particles was among the first signals 
measured at Relativistic Heavy Ion Collider (RHIC) in Brookhaven
\cite{star00}, and now there is plenty of data concerning the
centrality, the (pseudo)rapidity, and, especially, the transverse 
momentum dependence of the $v_2$ in gold-gold collisions at $\sqrt{s} 
= 130$ AGeV \cite{ell_fl_130} and at $\sqrt{s} = 200$ AGeV 
\cite{ell_fl_200}. Microscopic models based on string phenomenology 
and transport theory are able to reproduce, at least qualitatively, 
many features of the $v_2$ at ultrarelativistic energies 
\cite{ell_fl01,BlSt02,St04,LPX99,Sn00,CGG04,LK02,LK04}, 
however, the quantitative 
agreement with the data is often not so good. Particularly, magnitude 
of the distributions $v_2(\eta)$ or $v_2(p_t \geq 1.5$\,GeV/$c)$ 
appears to be too high. Does it mean that the effective EOS of hot 
and dense partonic-hadronic matter in microscopic models is too soft? 
What should be done to match the data? Before starting our study it 
is worth noting that the elliptic flow increases due to two main 
processes, namely, (i) elastic or inelastic rescattering of 
particles, and (ii) non-homogeneous absorption of particles in 
spatially asymmetric dense medium. The first reaction is responsible 
for the formation and development of the hydrodynamic flow, while 
the latter one should be considered as a non-hydro contribution.

Then, the microscopic calculations \cite{FO_ags,FO_sps,FO_sor,FO_urq}
show the absence of sharp freeze-out of particles in relativistic 
heavy-ion collisions. In contrast to assumptions of hydrodynamic model
\cite{Lan53}, the expanding fireball in microscopic models can be 
rather treated as a core consisting of still interacting hadrons, and
a halo, which contains particles already decoupled from the system. 
The order of the freeze-out of different species seems to be the same 
for energies ranging from AGS to RHIC: 1 - pions, 2 - kaons, 3 - 
lambdas, 4 - nucleons.
What are the consequences of the continuous freeze-out for the $v_2$ 
of these particles? What is the role of the rescattering and absorption
in the flow development? When (and where) the elliptic flow is formed?

\section{Model}
\label{sec2}

To answer these questions we employ the quark-gluon string model
(QGSM) \cite{qgsm1,qgsm2}, which is a microscopic model based on the
Gribov-Regge theory \cite{grt} of hadronic and nuclear interactions at 
high energies. The main advantage of the GRT is the fulfillment of
unitarity conditions in $s$- and $t$-channel for multiparticle
processes. On one hand, the Regge-pole exchange in $t$-channel 
determines two-particle amplitude at $s \rightarrow \infty$. On the 
other hand, Regge-pole exchanges fully determine multiparticle 
processes in $s$-channel, since each Regge-pole exchange in 
$t$-channel corresponds to a jet of hadrons with small transverse
momenta. Therefore, unitarity is fulfilled. In the QGSM the 
subprocesses with quark annihilation and quark exchange correspond to 
the so-called Reggeon exchange in two-particle amplitudes in the GRT, 
while the subprocesses with colour exchange are represented by the
one and more Pomeron exchanges in elastic amplitudes.
This scheme is closely related to that given by the partonic model,
where most of calculations imply somehow calculation of contributions
of diagrams arising in the GRT. More information concerning the link
between the GRT and quantum chromodynamics (QCD) can be found, e.g.,
in \cite{grt,Don02,BP02,Wer01} and references therein.       

Note also, that number of scattering partons (partonic density)
within hadrons or nuclei, colliding at ultrarelativistic energies,
increases with the rise of center-of-mass energy $\sqrt{s}$ as 
$s^{n[\alpha_P(0) -1]}$ with
$\alpha_P(0) > 1$ being the intercept of a Pomeranchuk pole.
This leads to the transition to Froissart regime corresponding to 
multiparton collision of black disks filled out by slow partons, i.e.,
the state which is a precursor of the Color Glass Condensate (see,
e.g., \cite{CGC01} and references therein).
The QGSM incorporates also the string
fragmentation, resonance formation, and hadronic rescattering. The
latter implies that the decay products of a string, - stable hadrons
and their resonances, - can further interact with other hadrons. Due
to uncertainty principle, secondary hadrons are allowed to interact 
again only after certain formation time, but hadrons containing the 
valence quarks of the colliding hadrons/nuclei can interact 
immediately with the reduced cross section $\sigma_{qN}$ taken from
the additive quark model. Angular and momentum distributions of 
secondaries are adjusted to available experimental data. In addition, 
the one-pion exchange model, detailed balance constructions, and 
isospin symmetry arguments are used in case of lacking the 
experimental information. The positions and momenta of nucleons inside
the colliding nuclei are Monte Carlo generated according to the 
Woods-Saxon density distribution and the Fermi momentum distribution,
respectively. The Pauli blocking is taken into account by excluding the
scattering into occupied final states.

The transverse motion of hadrons in the QGSM arises from different
sources: (i) primordial transverse momentum of the constituent quarks,
(ii) transverse momentum of (di)quark-anti(di)quark pairs acquired at
string breakup, (iii) the transverse Fermi motion of nucleons in
colliding nuclei, and (iv) rescattering of secondaries. Parameters of
the first two sources are fixed by comparison with hadronic data.
The Fermi motion changes the effective transverse distribution of
strings formed by the valence quarks and diquarks of the target and
projectile nucleons. Thus, the original strings are not
completely parallel to the beam axis. Further details of the QGSM
can be found elsewhere \cite{qgsm1,qgsm2}.    

\section{Development of elliptic flow}
\label{sec3}

\begin{figure}[hbt]
\centering{\
\epsfig{figure=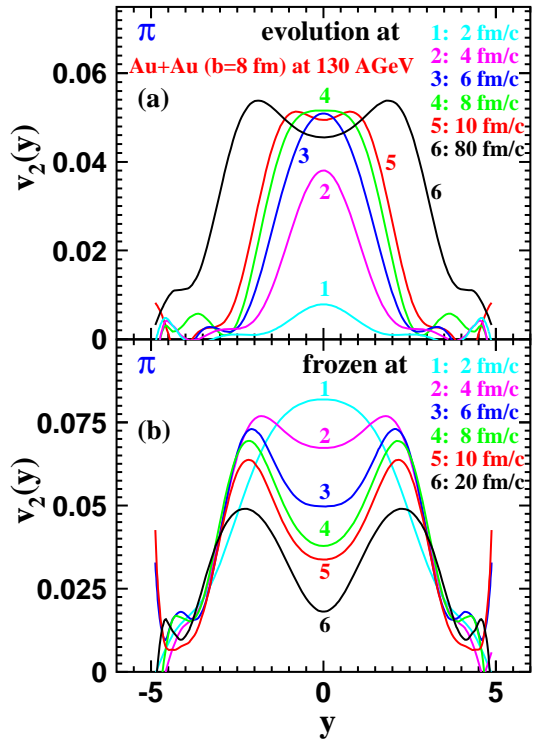,width=10.00cm}}
\caption{\label{fig1} Gold-gold collisions with $b=8$\,fm at 
$\sqrt{s}=130$ AGeV: (a) Evolution of elliptic flow of pions 
(snapshots are made at 2, 4, ... 80 fm/$c$), and (b) contributions to
the resulting $v_2^\pi (y)$ coming from pions frozen at 2, 4, ... 20 
fm/$c$, respectively.
}
\end{figure}

To investigate the development of the elliptic flow
ca. 20$\cdot 10^3$ gold-gold collisions with the impact parameter
$b = 8$\,fm were generated at $\sqrt{s} = 130$\, AGeV. According to
previous studies \cite{star00,ell_fl01} the elliptic flow of charged
particles is close to its maximum at this impact parameter, and the
multiplicity of secondaries is still quite high. The time evolutions
of the $v_2$ of pions and nucleons as functions of rapidity 
\beq \ds
v_2(y) =
\int_{0}^{p_t^{max}} \cos(2\phi) \frac{d^2 N}{d y d p_t}
d p_t \left/ \int_{0}^{p_t^{max}}
\frac{d^2 N}{d y d p_t} d p_t \right.
\label{eq4}
\eeq
are displayed in Fig.~\ref{fig1}(a). Here the snapshots of the $v_2$ 
profile are taken at 2 fm/$c$, 4 fm/$c$, $\ldots$ 10 fm/$c$, and at 
the time of thermal freeze-out. This means
that the hadronic composition was frozen at certain time $t = t_i$,
when all interactions were switched off and particles were propagated
freely. To avoid ambiguities, resonances were allowed to decay
according to their branching ratios. Surprisingly, at $t = 2$ fm/$c$
elliptic flow of pions is weak. The flow continuously increases and
reaches its maximum value $v_2^\pi (y=0) \approx 5\%$ already at 
$t = 6$ fm/$c$. From this time the elliptic flow does not increase 
anymore.
Instead, it becomes broader and develops a two-hump structure with a
relatively weak dip at midrapidity. The flow seems to continue
development till the late stages of the system evolution. Bearing in
mind the absence of sharp freeze-out of particles in model 
calculations, it is important to study the contribution of the 
survived particles to the resulting elliptic flow. The corresponding
rapidity distributions presented in Fig.~\ref{fig1}(b) reveal the
peculiar feature: the $v_2$ of pions, which are frozen already at 
$t=2$ fm/$c$, is the {\bf strongest} among the fractions of the flow
carried by pions decoupled from the fireball later on. The later the
pions are frozen, the weaker their flow. The two-hump structure of the 
signal develops here as well, but the widths of all rapidity 
distributions are the same. For pions frozen after $t = 6$ fm/$c$ the
dip at midrapidity is seen quite distinctly.
From here one can conclude that the strong elliptic anisotropy of 
pions, which left the system early, is caused by the absorption of the
pion component in the squeeze-out direction.  

\begin{figure}[hbt]
\centering{\
\epsfig{figure=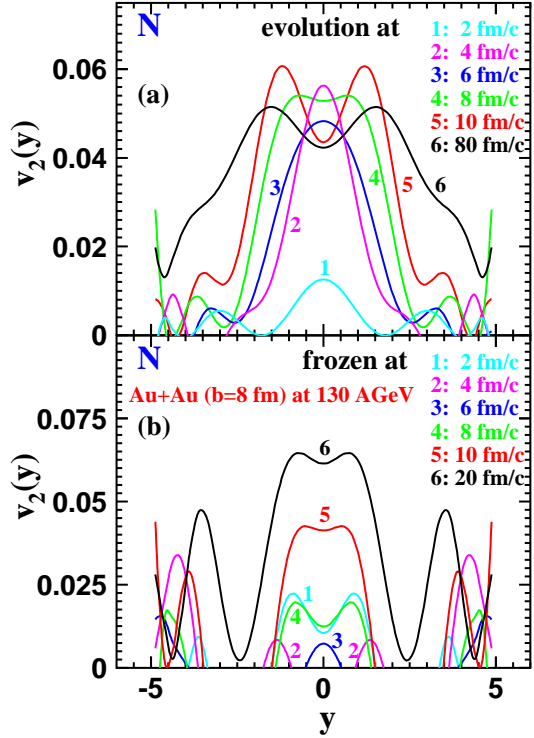,width=10.00cm}}
\caption{\label{fig2} The same as Fig.\protect\ref{fig1} but for 
the elliptic flow of nucleons.
}
\end{figure}

For nucleons the evolution picture of the $v_2(y)$, shown in 
Fig.~\ref{fig2}(a), is similar to that for pions. The flow is quite
weak at $t = 2$\,fm/$c$, then it increases and gets a full strength at
midrapidity between 8\,fm/$c$ and 10\,fm/$c$, i.e. later than the 
elliptic flow of pions. Similarly to $v_2^\pi (y)$, it develops a
two-hump structure, but the humps tend to dissolve at late stages of
system evolution. In contrast to this behavior, the freeze-out 
decomposition picture of $v_2^N (y)$, presented in Fig.~\ref{fig2}(b),
does not show monotonic tendency within first 8\,fm/$c$ of the 
reaction: The flow of nucleons frozen at 2\,fm/$c$ is identical to
that of nucleons frozen at 8\,fm/$c$, whereas nucleons decoupled from
the system between 2\,fm/$c$ and 8\,fm/$c$ almost do not contribute
to the resulting elliptic flow. Nucleons, which are decoupled after
8\,fm/$c$, have significant anisotropy in the momentum space, and
the time development of flow in the nucleon sector is opposite to 
that in the pion one. Namely, the later the nucleons are frozen, the
{\bf stronger} their elliptic flow.

\begin{figure}[hbt]
\centering{\
\epsfig{figure=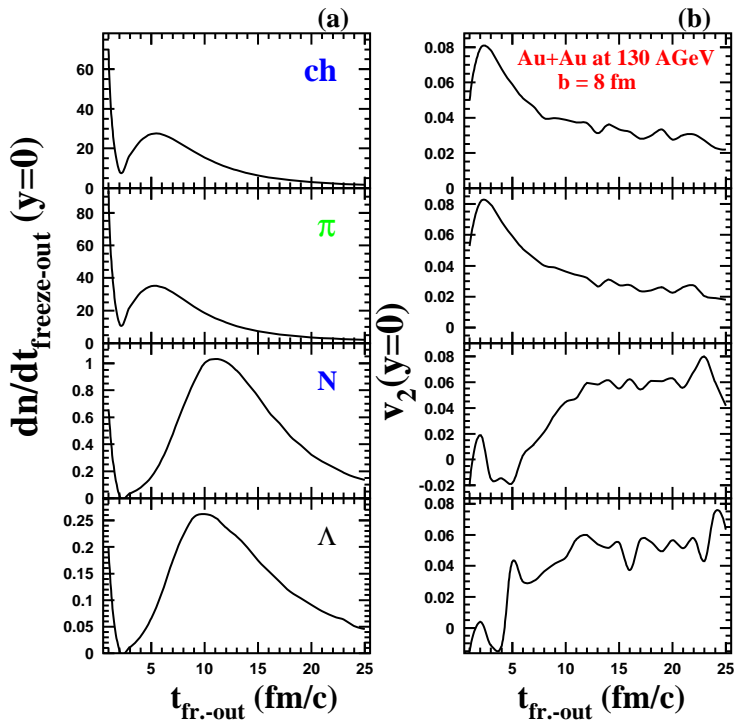,width=10.00cm}}
\caption{\label{fig3} 
(a) $dN/dt$ distribution of $n_{ch}, \pi, N, \Lambda$ over
the time of their last interaction, and (b)
elliptic flow of these particles for Au+Au collisions with $b=8$\,fm
at $\sqrt{s}=130$ AGeV.
}
\end{figure}

To clarify the role of particle freeze-out for the formation and
evolution of their elliptic flow,
the $d N /d t$ distributions of $n_{ch},\ \pi,\ N,\ {\rm and}\ 
\Lambda$, which are decoupled from the system after the last elastic 
or inelastic collision, are depicted in Fig.~\ref{fig3}(a). Here two
features should be mentioned. Firstly, a substantial part of hadrons 
leave the fireball immediately after their production within the first
two fm/$c$, in stark contrast with heavy-ion reactions at AGS 
\cite{FO_ags} and at SPS \cite{FO_sps} energies. Secondly, compared 
to lower energies, the particle distributions at RHIC are peaking at 
earlier times. For instance, the pion distribution has maximum at 
$t_{max}^\pi = 5$-6 fm/$c$, while the distributions of baryons are 
broader due to large number of rescatterings shifting their maxima to 
later times, $t_{max}^B = 10$-12 fm/$c$ (cf. $t_{max}^\pi \approx 10$ 
fm/$c$ and $t_{max}^B \approx 18$ fm/$c$ at both AGS and SPS energies). 

Elliptic flow carried by these hadronic species is presented in 
Fig.~\ref{fig3}(b). The baryonic and mesonic components are
completely different: pions emitted from the surface of the expanding
fireball within the first few fm/$c$ carry the strongest flow, while
later on the flow of pions is significantly reduced. In contrast to 
pions, the baryon fraction acquires stronger elliptic flow during the
subsequent rescatterings, thus developing the hydro-like flow.
The saturation of the flow at the late stages can be explained
by the lack of rescattering as the expanding system becomes
more dilute, and by the restoration of the symmetry of particle
momentum distributions in the transverse plane.

\begin{figure}[hbt]
\centering{\
\epsfig{figure=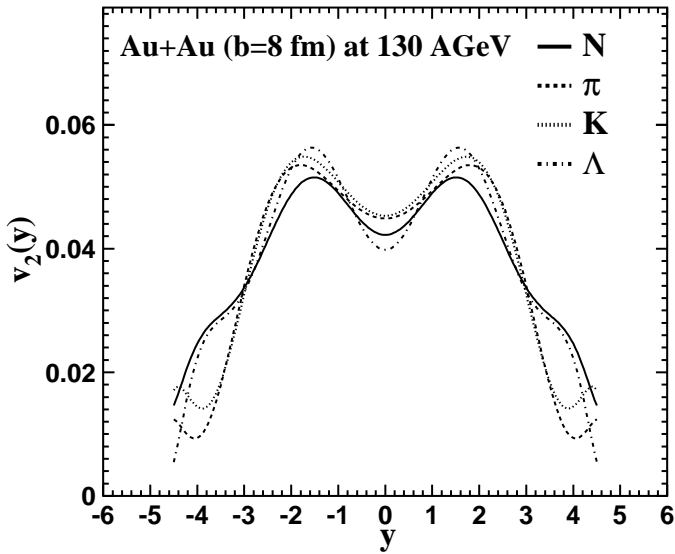,width=10.00cm}}
\caption{\label{fig4} 
Rapidity distribution of the resulting elliptic flow of
nucleons (full line), pions (dashed line), kaons (dotted line), 
and lambdas (dash-dotted line) in Au+Au collisions with $b=8$\,fm at 
$\sqrt{s}=130$ AGeV.
}
\end{figure}

The flow has a multicomponent structure. In addition to hydrodynamic
flow there are other components of the $v_2$ caused mainly by the
particle splash from the surface area during the initial phase of
nuclear collision, and by the non-uniform absorption of hadrons in
spatially asymmetric dense matter. For instance, the anisotropic 
absorption of jets in $x$- and $y$-directions will contribute to the
development of the elliptic flow of particles \cite{LoSn02} with high 
transverse momenta. 
Hence, one may expect that the resulting elliptic flows of
e.g. baryons and mesons obtained after the particle freeze-out are
different. Figure \ref{fig4} shows the rapidity dependence of the
$v_2$ of most abundant particle species in Au+Au collisions at RHIC
energies, namely pions, kaons, nucleons, and lambdas. Surprisingly,
these distributions are almost indistinguishable: the widths and
magnitudes of the signals are the same. To check the validity of the
model calculations one has to perform more complex analysis of 
experimental data e.g. by studying the $v_2(y)$ distributions in
different $p_T$-intervals.  
\begin{figure}[hbt]
\centering{\
\epsfig{figure=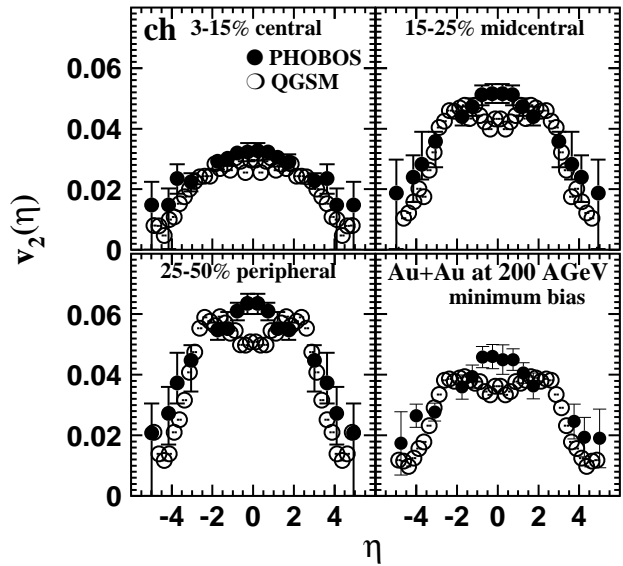,width=10.00cm}}
\caption{\label{fig5} 
$v_2(\eta)$ distribution of charged particles in Au+Au collisions at
$\sqrt{s}=200$ AGeV for (a) $\sigma/\sigma_{geo} = 0-15\%$,
(b) $\sigma/\sigma_{geo} = 15-25\%$, (c) $\sigma/\sigma_{geo} = 
25-50\%$, and (d) minimum bias events. Full symbols represent data 
\protect \cite{phob_v2_200}, open symbols denote model calculations.
 }
\end{figure}
Other model predictions for the rapidity distribution of the elliptic 
flow of hadrons at both RHIC energies, 130 AGeV and 200 AGeV, can be 
found in \cite{ell_fl01,sqm01,ppnp04,Bur04}. 
Recently, PHOBOS collaboration 
presented the data on the $v_2(\eta)$ distributions of charged 
particles, produced in Au+Au collisions at 200 AGeV, in different
centrality bins \cite{phob_v2_200}. Model calculations are plotted
onto the experimental results in Fig. \ref{fig5}. Here the elliptic
flow of charged particles for the collisions with centralities 
0-15\% ($0\, fm \leq b < 2.3\, fm$), 15-25\% ($2.3\, fm \leq b < 
6.5\, fm$), and 25-50\% ($6.5\, fm \leq b < 9.2\, fm$) is 
presented together with the resulting flow in minimum bias events.
We see that the model reproduces the measured signal pretty well,
including the nearly flat distribution at $|\eta| \leq 2$ and quick
fall of the elliptic flow at $|\eta| \geq 2$. The only discrepancy 
appears to arise in semiperipheral collisions at midrapidity range,
where the data are approximately 15\% above the model results.
This problem can be fixed by (i) the fine tuning of model parameters,
and (ii) proper analysis of the experimental procedure of data 
handling (for the analysis and comparison of several methods applied
to restore the anisotropic flow see \cite{BO_01}).    
 
\section{Conclusions}
\label{sec4}

In summary, the features of the formation and development of elliptic 
flow in the microscopic quark-gluon string model can be stated as 
follows. First of all, there is no one-to-one correspondence between 
the apparent elliptic flow and the contribution to the final flow 
coming from the ``survived" fraction of particles. 
For instance, apparent elliptic flow of pions at $t =2$ fm/$c$ is 
weak, but pions which are already decoupled from the system at this 
moment have the strongest elliptic anisotropy caused by the absorption 
of the pion component in the squeeze-out direction.
Elliptic flow of hadrons is formed not only during the first few 
fm/$c$, but also during the whole evolution of the system because of
continuous freeze-out of particles. Secondly, time evolutions of the 
mesonic flow and baryonic flow are quite different. Baryons, except of
the fraction emitted within first two fm/$c$ after the collision, are
getting the stronger $v_2$ in the course of secondary interactions.
Pions in average experience much less elastic collisions per particle
because of their instant absorption and production in dense medium.
As the substance becomes more dilute, their elliptic flow decreases.
Freeze-out dynamics for baryons and mesons is also different, 
therefore, development of particle collective flow should not be
studied independently of the freeze-out picture.   
The general trend in particle flow formation in microscopic models at 
ultrarelativistic energies is that the earlier mesons
are frozen, the weaker their elliptic flow. In contrast, baryons 
frozen at the end of the system evolution have stronger $v_2$.

{\it Acknowledgments.}
Fruitful discussions with L. Csernai, R. Lacey, S. Panitkin, and
S. Voloshin are gratefully acknowledged.
This work was supported by the Norwegian Research Council (NFR) and by 
the Bundesministerium f{\"u}r Bildung und Forschung under contract 
06T\"U986.

\end{document}